\documentstyle[12pt,epsf,axodraw]{article}
\title{$\tau$ polarization and Randall-Sundrum scenario
at $e^+e^-$ colliders}

\author{Namit Mahajan\thanks{E--mail : nm@ducos.ernet.in, 
nmahajan@physics.du.ac.in} \\
	{\em Department of Physics and Astrophysics,} \\
	 {\em University of Delhi, Delhi-110 007, India.}}

\begin{document}
\maketitle
\begin{abstract}
 We study polarized cross-sections and forward-backward asymmetry
 for the process $e^+e-\rightarrow 
\tau^+\tau^-$ in the stabilized Randall-Sundrum scenario. It is 
shown that there is substantial deviation from the Standard Model
predictions, both in terms of the actual numerical values and angular
distributions.  \\ \\
\end{abstract}
\indent The hierarchy between the electroweak and the Planck scale
remains to be a mystry till date. Many possible solutions have been 
proposed. The idea that the fundamental scale of gravity is not as large
as the Planck scale, ${\mathcal{O}}(10^{-19}$ GeV), but could be
as low as ${\mathcal{O}}$(TeV) has invited considerable attention.
The basic idea behind such a proposal \cite{nima} is the existence of n
 compact extra spatial
dimensions. The Standard Model (SM) fields all lie on a 3-brane
while gravity is free to propagate in the entire bulk. The four dimensional
Planck scale, $M_{Pl}$, gets related to the fundamental scale of gravity,
$M_*$, and the 
volume, $V_n$, of the n extra dimensions as
\begin{equation}
M_{Pl}^2~\sim ~V_nM_*^{n+2}
\end{equation}
In this picture the bulk space-time is a direct product of the four-dimenisoal
Minkowski space and the extra dimensions. To attain the relevant numbers,
the volume $V_n$ should be large and this in turn introduces a new 
hierarchy between the electroweak scale and the inverse of the size of 
these extra dimensions. \\
\indent An alternative scenario due to Randall and Sundrum \cite{rs}
assumes the existence of two 3-branes which define the ends of the 
world in the context of a five dimensional bulk space-time.
The bulk geometry in this case, in contrast to the previous one, is 
non-factorizable and the hierarchy is generated by the
warp factor which is an exponential function of the inter-brane distance.
 The same warp factor relates the induced metrics on the two branes.
The question of stability of this inter-brane distance still remains
an important issue of concern. It was shown by Goldberger and Wise \cite{gw}
that the inclusion of a bulk scalar field minimally coupled to gravity
does indeed admit a stable solution. \\
\indent In the present work, we assume that such a stabilizing mechanism
is at play. The bulk space-time is a slice of $AdS_5$ and the metric is
\begin{equation}
ds^2 = e^{-2kr_c\vert y\vert}\eta_{\mu\nu}dx^{\mu}dx^{\nu} - r_c^2dy^2
\end{equation}
where $x^{\mu}$ are the usual four dimensional coordinates and $y \in 
[-\pi,\pi]$ parametrizes an $S^1/Z_2$ orbifold. The quantity $r_c$ is the 
radius and $k^{-1}$ is the curvature of this $AdS_5$ space-time. The two 
branes lie at $y=0$ and $y=\pi$ and have equal and opposite tensions.\\
\indent The fluctuations about the flat metric, $\eta_{\mu\nu}$, contain
both the four dimenional massless graviton and its massive Kaluza-Klein
(KK) excitations.
 On the other hand quantum fluctuations of the inter-brane 
distance manifest themselves in the form of a scalar particle describing
the modulus field. Inclusion of the bulk scalar field as in \cite{gw}
provides a nontrivial potential for the modulus field, thereby stabilizing
the inter-brane distance. 
 It is quite clear that to an observer sitting on the 
$y=\pi$ brane, a field with fundamental mass scale $m_0$ will appear to
have an effective mass $m=m_0e^{-kr_c\pi}$. Therefore, with $kr_c\sim 12$,
TeV scales are easily generated on the $y=\pi$ brane and no other hierarhcy
is introduced.\\
\indent The couplings of SM fields with the massless and massive gravitons,
and the modulus field can be easily found out \cite{rs,gw,csaba,hewett}.
On the $y=\pi$ brane, we have after compactification, to the lowest order
\begin{equation}
{\mathcal{L}}_{int} = -\frac{1}{M_{Pl}}T^{\alpha\beta}(x)h^{(0)}_{\alpha\beta}
- \frac{1}{\Lambda_{\pi}}T^{\alpha\beta}(x)\sum_{n=1}^{\infty}
h^{(n)}_{\alpha\beta} - \frac{\varphi}{\langle\varphi\rangle}
T^{\alpha}_{\alpha}
\end{equation}
Here $h^{(n)}_{\alpha\beta}$ are the massive gravitons and $\varphi$ is the
modulus field with coupling strengths
$\Lambda_{\pi}$ and $\langle\varphi\rangle$ (both ${\cal{O}}$(TeV))
 respectively. 
$h_{\alpha\beta}^{(0)}$ is the massless four dimensional graviton and 
$T^{\alpha\beta}$
is the energy-momentum tensor for the SM fields.
From this equation it is obvious to see that to the lowest order, 
the modulus field couples to the SM fileds like the Higgs boson. In particular,
the modulus field couples to the fermions via the fermion mass.\\
\indent In the present work, we consider tau pair production at the 
$e^+e^-$ linear colliders. The presence of the extra dimension leads to
additional contributions to the process $e^+e-\rightarrow \tau^+\tau^-$
through the gravitational interactions induced by the gravitons and
the modulus field. As can be seen from the interaction Lagrangian, the
usual four dimensional massless graviton couples with the ordinary 
gravitational strength and thus its contribution is rather negligible.
Also, the coupling of the modulus field with the fermions is proportional
to fermion mass. Therefore, in the present case (with $m_e=0$), the 
contribution due to the exchange of this modulus field is also neglected.
We are thus left with an additional contribution from the massive graviton
modes only. As is known \cite{csaba,hewett} that in Randall-Sundrum scenario,
 the mass spectrum of
these graviton modes is discrete  with the first massive graviton having
a mass ${\cal{O}}$(TeV) and the successive modes are expected to be separated 
by ${\cal{O}}$(TeV). This leads to considering only the first massive
graviton exchange contribution.\\ \\

\indent The process $e^+e-\rightarrow \tau^+\tau^-$ receives contribution
only from the s-channel diagram mediated by the massive spin-2 gravitons,
apart from the SM contributions mediated by the photon and the Z-boson
(Figure1.)  

\vskip 1cm
\begin{center}
 \begin{figure}[htb]
\vspace*{-8ex}
\hspace*{2em}
\begin{tabbing}
\hskip 2cm
\begin{picture}(155,120)(-5.0,-20)
\Line(0,45)(45,0)
	\Text(15,50)[c]{$e^-(p_1)$}
\ArrowLine(20,40)(30,30)
\Line(0,-45)(45,0)
	\Text(15,-50)[c]{$e^+(p_2)$}
\ArrowLine(20,-40)(30,-30)
\Photon(45,0)(90,0){3}{6}
	\Text(65,15)[c]{$\gamma /Z$}
\Line(90,0)(135,45)
	\Text(120,50)[c]{$\tau^-(k_1)$}
\ArrowLine(105,30)(115,40)
\Line(90,0)(135,-45)
	\Text(120,-50)[c]{$\tau^+(k_2)$}
\ArrowLine(105,-30)(115,-40)
\end{picture}
\hskip 1.5cm
\begin{picture}(155,120)(-5.0,-20)
\Line(0,45)(45,0)
	\Text(15,50)[c]{$e^-(p_1)$}
\ArrowLine(20,40)(30,30)
\Line(0,-45)(45,0)
	\Text(15,-50)[c]{$e^+(p_2)$}
\ArrowLine(20,-40)(30,-30)
\Gluon(45,0)(90,0){3}{6}
\Photon(45,0)(90,0){3}{6}
	\Text(65,15)[c]{$h_{\mu\nu}$}
\Line(90,0)(135,45)
	\Text(120,50)[c]{$\tau^-(k_1)$}
\ArrowLine(105,30)(115,40)
\Line(90,0)(135,-45)
	\Text(120,-50)[c]{$\tau^+(k_2)$}
\ArrowLine(105,-30)(115,-40)
\end{picture}
\end{tabbing}
\vskip 2cm
\caption{\em The s-channel contributions due to photon, Z and 
massive gravitons.}
\end{figure}
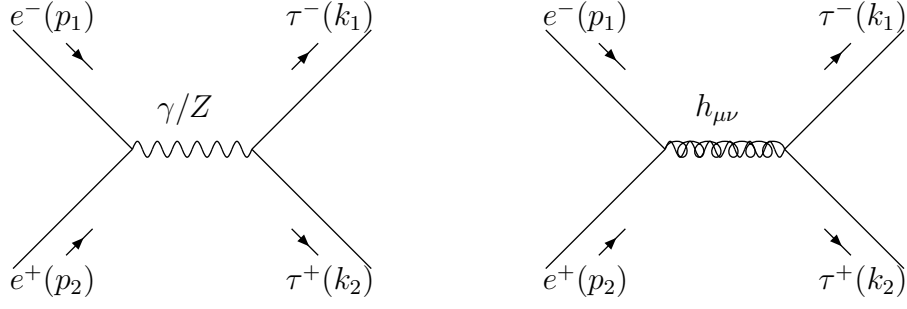
\end{center}
\indent In SM, the scattering process takes place through the exchange
of $\gamma /Z$, both of which are spin-1 particles. The situation
becomes more interesting with the presence of a spin-2 graviton. The spin-2 
nature of the graviton is responsible for significant deviation 
from the SM results. Moreover, SM being a renormalizable theory gives 
contributions that are well behaved at high energies. But the 
non-renormalizable gravitational interactions, in addition to SM interactions,
lead to amplitudes not well behaved at higher energies. 
Therefore, any significant deviation from the SM predictions at high energies
may serve as an indication of the possible existence of such interactions.
The Next Linear Colliders, thus can be a good testing place for such 
theories.\\
\indent For the process 
\[
e^+(p_2)e^-(p_1)\rightarrow \tau^+(k_2)\tau^-(k_1)\, ,
\]
the scattering amplitude is
\begin{equation}
{\mathcal{M}} = {\mathcal{M}}_{\gamma} + {\mathcal{M}}_Z + {\mathcal{M}}_G
\end{equation}
where
\begin{equation}
{\mathcal{M}}_{\gamma} = \Bigg(\frac{\imath e^2}{s}\Bigg)~\bar{u}(k_1)
\gamma_{\mu}v(k_2)~ \bar{v}(p_2)\gamma^{\mu}u(p_1)
\end{equation}
$e$ is the electromagnetic charge and $(p_1+p_2)^2=s=(k_1+k_2)^2$.

\begin{equation}
{\mathcal{M}}_Z = \Bigg(\frac{\imath g^2}{16 c_w^2}\Bigg)
\frac{1}{s-M_Z^2}~\bar{u}(k_1)
\gamma_{\mu}(a-\gamma_5)v(k_2)~\bar{v}(p_2)\gamma^{\mu}(a-\gamma_5)u(p_1)
\end{equation}
Here $g$ is the weak coupling, $c_w$ is the cosine of Weinberg angle
and the parameter $a=-1+4\sin^2\theta_w$ is nothing but the vector 
coupling of the fermions to the Z-boson.
\begin{eqnarray}
{\mathcal{M}}_G &=& \Bigg(-\frac{\imath}{\Lambda_{\pi}^2}\Bigg)
\frac{1}{s-M_G^2}~\Bigg[(k^{\prime}\cdot p^{\prime})\bar{u}(k_1)
\gamma_{\mu}v(k_2)~\bar{v}(p_2)\gamma^{\mu}u(p_1) \\ \nonumber
&+& \bar{u}(k_1)\not{p^{\prime}}v(k_2)~\bar{v}(p_2)\not{k^{\prime}}
u(p_1)\Bigg]
\end{eqnarray}
In the above expression
\[
p^{\prime} = p_1 - p_2 \hskip 1.5cm  k^{\prime} = k_1 - k_2
\]
\indent From these expressins one can easily compute the polarized 
amplitudes for different possible combinations. The electrons are taken
to be massless. Also, at the energy range at which the linear collider
is expected to operate, the $\tau$'s can also be treated as massless.
 We follow the notation and
convention as in \cite{kleiss} to compute these amplitudes. In the 
helicity basis, only the amplitudes with opposite (off-diagonal) helicity 
for both the 
pairs of external particles are non-vanishing. In what follows below,
we label a right handed particle $\psi_R$ as $\psi(+)$ and similarly
a left handed particle $\psi_L$ as $\psi(-)$ \footnote{chirality equals
helicity for massless particles while it is equal to the negative of 
helicity for massless antiparticles}. With these conventions, we have 
the following non-vanishing amplitude squares (as functions of invariants
$s$ and $(p_1-k_1)^2=t=(k_2-p_2)^2$): 
\begin{equation}
\vert{\mathcal{M}}(+-;+-)\vert^2 = 4(s+t)^2\Bigg[\frac{4\pi\alpha_{em}}{s} + 
\Bigg(\frac{g^2}{16c_w^2}\Bigg)\frac{(a-1)^2}{s-M_Z^2} + 
\frac{s+4t}{\Lambda_{\pi}^2(s-M_G^2)}\Bigg]^2
\end{equation}

\begin{eqnarray}
\vert{\mathcal{M}}(+-;-+)\vert^2 &=& 
\vert{\mathcal{M}}(-+;+-)\vert^2 \\ \nonumber
&=& 4{t}^2 \Bigg[\frac{4\pi\alpha_{em}}{s} + 
\Bigg(\frac{g^2}{16c_w^2}\Bigg)\frac{(a^2-1)}{s-M_Z^2} + 
\frac{3s+4t}{\Lambda_{\pi}^2(s-M_G^2)}\Bigg]^2
\end{eqnarray}

\begin{equation}
\vert{\mathcal{M}}(-+;-+)\vert^2 = 4(s+t)^2\Bigg[\frac{4\pi\alpha_{em}}{s} + 
\Bigg(\frac{g^2}{16c_w^2}\Bigg)\frac{(a+1)^2}{s-M_Z^2} + 
\frac{s+4t}{\Lambda_{\pi}^2(s-M_G^2)}\Bigg]^2
\end{equation}
\indent Let $\theta$ be the angle between the incoming $e^-$ and the 
outgoing $\tau^-$. The forward-backward asymmetry 
\[
{\mathcal{A}}_{FB} = \frac{\int_0^1\frac{d\sigma}{d\cos\theta}d\cos\theta - 
 \int_{-1}^0\frac{d\sigma}{d\cos\theta}d\cos\theta}
{\int_0^1\frac{d\sigma}{d\cos\theta}d\cos\theta + 
 \int_{-1}^0\frac{d\sigma}{d\cos\theta}d\cos\theta}
\]
for different cases is summarised in the following table :

\begin{table}[ht]
\begin{center}
\begin{tabular}{|l|l|r|l|l|l|}\hline
$\sigma(+-;+-)$&$\sqrt{s}=500$GeV &SM=0.75&SM+RS=-0.210615\\ \cline{2-3}
\cline{3-4}
&$\sqrt{s}=1000$GeV &SM=0.75&SM+RS=0.119273 \\ \hline
$\sigma(+-;-+)$&$\sqrt{s}=500$GeV &SM=-0.75&SM+RS=-0.149843\\ \cline{2-3}
\cline{3-4}
&$\sqrt{s}=1000$GeV &SM=-0.75&SM+RS=0.0485942 \\ \hline
$\sigma(-+;-+)$&$\sqrt{s}=500$GeV &SM=0.75&SM+RS=-0.201902\\ \cline{2-3}
\cline{3-4}
&$\sqrt{s}=1000$GeV &SM=0.75&SM+RS=0.109656 \\ \hline
\end{tabular}
\caption{The ${\mathcal{A}}_{FB}$ values for $\sqrt{s}=500,~1000$ GeV
($M_G=600$ GeV, $\Lambda_{\pi}=1000$ GeV). $SM$ and $SM+RS$ are the SM and
combined results.}
\end{center}
\end{table}
It is quite evident that there is a large deviation from the SM values
when one has extra contribution from the massive graviton states, making 
such a measurement a possible testing ground for such theories.\\
\indent Also, the spin-2 nature of the gravitons is responsible for 
significant change in the angular distribution of various differential
cross-sections as compared to SM case where both the mediating particles 
are spin-1 objects. This is another feature that can be used to 
establish the existence of such theories and put meaningful constraints on 
the parameters involved.\\ 
\vskip 2.0cm
\begin{figure}[ht]
\vspace*{-1.0cm}
\centerline{
\epsfxsize=9.0cm\epsfysize=8.0cm
                     \epsfbox{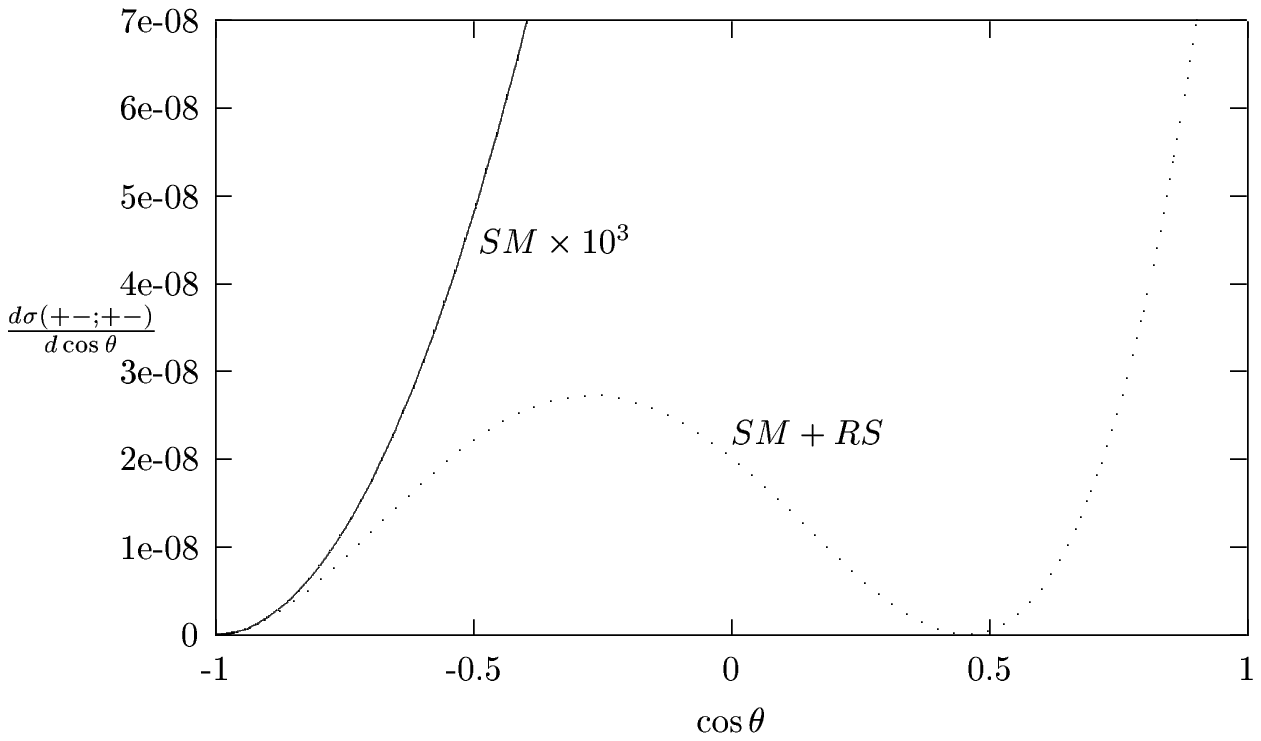}
}
\caption{\em Differential cross-section for the process $e^-_Re^+_L
\rightarrow \tau^-_R\tau^+_L$ with respect to the scattering
angle of $\tau^-$ relative to $e^-$ at $\sqrt{s}=1000$ GeV, $M_G=600$ GeV
and $\Lambda_{\pi}=1000$ GeV.}
\end{figure}
\vskip 3cm
\begin{figure}[ht]
\vspace*{-3.5cm}
\centerline{
\epsfxsize=9.0cm\epsfysize=8.0cm
                     \epsfbox{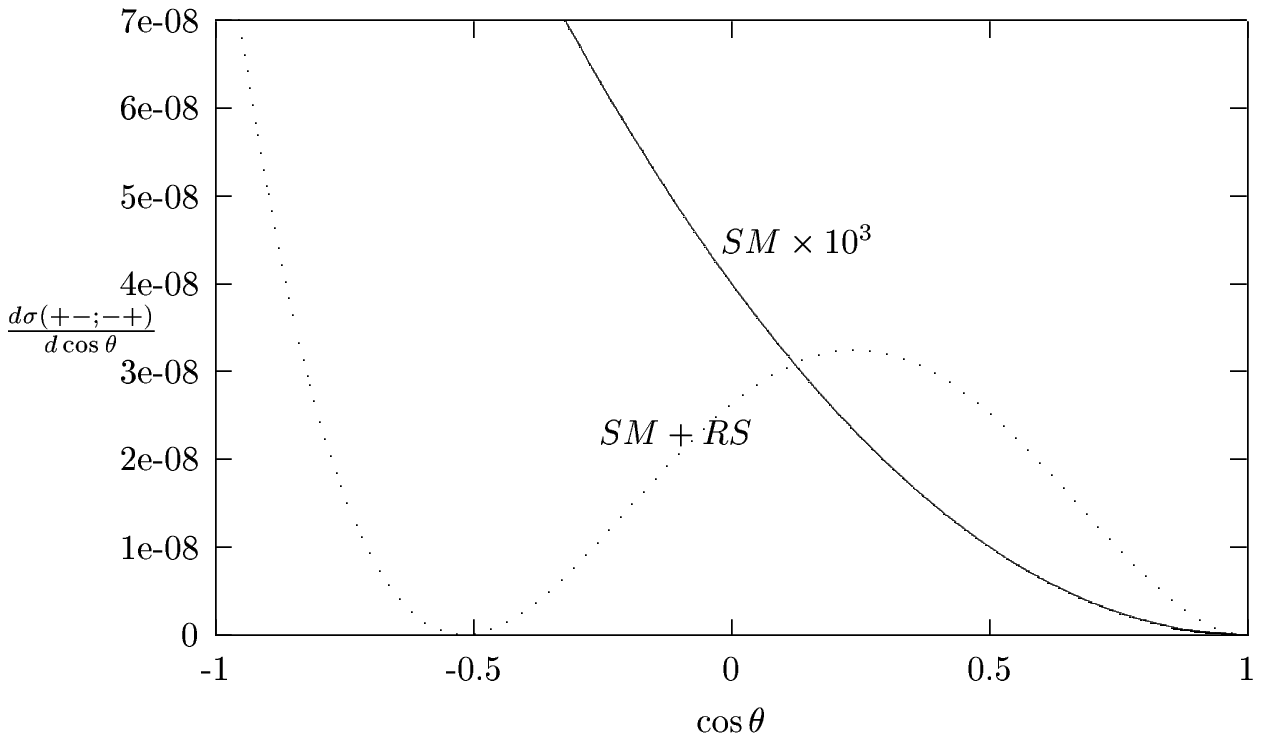}
}
\caption{\em Differential cross-section for the processes $e^-_{R,L}e^+_{L,R}
\rightarrow \tau^-_{L,R}\tau^+_{R,L}$ with respect to the scattering
angle of $\tau^-$ relative to $e^-$ at $\sqrt{s}=1000$ GeV, $M_G=600$ GeV
and $\Lambda_{\pi}=1000$ GeV.}
\end{figure}
\vskip 5.0cm
\begin{figure}[ht]
\vspace*{-0.15cm}
\centerline{
\epsfxsize=9.0cm\epsfysize=8.0cm
                     \epsfbox{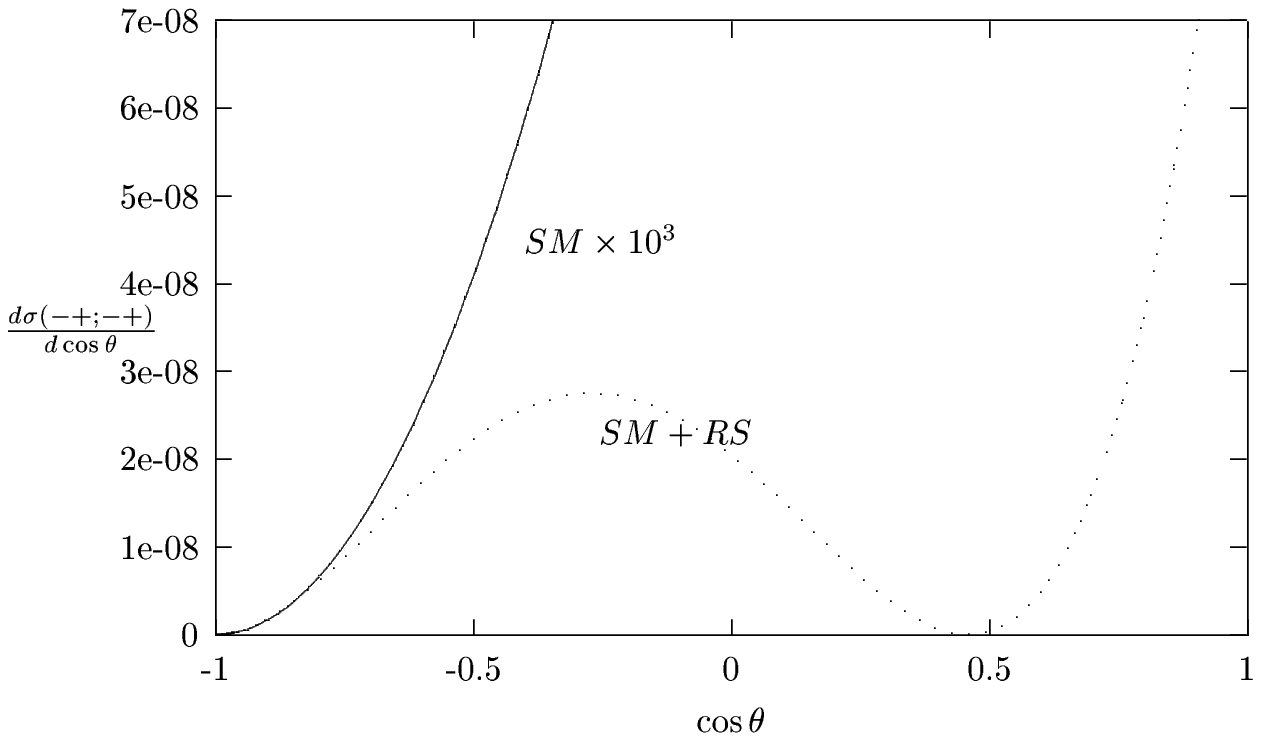}
}
\caption{\em Differential cross-section for the process $e^-_Le^+_R
\rightarrow \tau^-_L\tau^+_R$ with respect to the scattering
angle of $\tau^-$ relative to $e^-$ at $\sqrt{s}=1000$ GeV, $M_G=600$ GeV
and $\Lambda_{\pi}=1000$ GeV.}
\end{figure}
\indent In the Figs. (2)-(4), the pure SM contribution
has been scaled by three orders of magnitude and is denoted by
$SM\times 10^3$ while $SM+RS$ denotes the combined contribution
of SM and gravitons. The later is far above the former and also 
the distribution is quite different as expected.\\
\indent In whatever has been discussed above, we have assumed $100\%$
polarization of the initial beams. But in practice this may not be the case.
It is worth emphasizing that the same expressions with obvious modifications
can be used to accomodate the partial/zero degree of polarization of 
the initial beams. \\
\indent Further, we can calculate the final state polarization asymmetry
\cite{hagiwara} 
\[
{\cal{P}}_{\tau^-} = \frac{\sigma_{\tau^-_R} - \sigma_{\tau^-_L}}
 {\sigma_{\tau^-_R} + \sigma_{\tau^-_L}}
\]   
Again, we expect it to be different from the SM value. In the case at hand,
it is the graviton and Z-boson cross contribution that gives the 
additional contribution to the asymmetry. The results are presented in
the following table:

\begin{table}[ht]
\begin{center}
\begin{tabular}{|l|l|r|l|l|l|}\hline
$\frac{\sigma(+-;+-)-\sigma(+-;-+)}{\sigma(+-;+-)+\sigma(+-;-+)}$&
$\sqrt{s}=500$GeV &SM=0.666&SM+RS=0.038\\
 \cline{2-3}\cline{3-4}
&$\sqrt{s}=1000$GeV &SM=0.654&SM+RS=0.0052 \\
 \hline
$\frac{\sigma(-+;+-)-\sigma(-+;-+)}{\sigma(-+;+-)+\sigma(-+;-+)}$&
$\sqrt{s}=500$GeV &SM=0.6183&SM+RS=0.0311\\
 \cline{2-3}\cline{3-4}
&$\sqrt{s}=1000$GeV &SM=0.6056&SM+RS=0.0419 \\
 \hline
\end{tabular}
\caption{The final state polarization asymmetry for $\sqrt{s}=500,~1000$ GeV
($M_G=600$ GeV, $\Lambda_{\pi}=1000$ GeV).}
\end{center}
\end{table}
\indent The $\tau$ polarization measurement (through the study of correlations
between the decay products) \cite{tsai} can provide further useful
hints about the existence of physics beyond SM. It is then straight forward
to use the above relations and study correlations between the decay products
of $\tau^+$ and $\tau^-$. For example in the present case, $\pi^-$, 
from the decay channel 
$\tau^-\rightarrow \pi^-\nu_{\tau}$, will be emitted in the $\tau$-spin 
direction in the rest frame of $\tau^-$ while $\pi^+$ from 
$\tau^+\rightarrow \pi^+\bar{\nu_{\tau}}$ emitted in the direction
opposite to the $\tau^+$ spin. Thus, an appropriate Lorentz transformation
to the Lab frame can indicate at the expected angular distribution 
of $\pi^{\pm}$
and provide information about deviation from SM.\\ \\
\indent In conclusion, we can say that $\tau$ polarization can give
useful hints on the possible existence of Randall-Sundrum type scenarios
and can be used to constrain the parameters of such theories
 \footnote{one expects similar results in the case of the scenario 
proposed by Arkani-Hamed etal \cite{nima} where instead of a single 
graviton exchange, one has a tower of massive gravitons till the effective
Planck/string scale, $M_*$}.


\begin{section}*{Acknowledgements}
The author would like to thank University Grants Commission,
 India for fellowship. 
\end{section} 

\end{document}